\documentclass{article}

\usepackage{arxiv}

\usepackage[utf8]{inputenc} 
\usepackage[T1]{fontenc}    
\usepackage{hyperref}       
\usepackage{url}            
\usepackage{booktabs}       
\usepackage{amsfonts}       
\usepackage{nicefrac}       
\usepackage{microtype}      
\usepackage{lipsum}		
\usepackage{graphicx}
\usepackage{natbib}
\usepackage{doi}

\title{The SpacePy space science package at 12 years}

\author{ \href{https://orcid.org/0000-0001-6286-5809}{\includegraphics[scale=0.06]{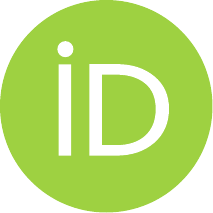}\hspace{1mm}Jonathan T. Niehof}\\
	Space Science Center\\
	University of New Hampshire\\
	Durham, NH \\
	\texttt{jonathan.niehof@unh.edu} \\
	\And
	\href{https://orcid.org/0000-0001-8520-0199}{\includegraphics[scale=0.06]{orcid.pdf}\hspace{1mm}Steven K. Morley} \\
	Space Science and Applications\\
	Los Alamos National Laboratory\\
	Los Alamos, NM\\
	\\
	\And
	\href{https://orcid.org/0000-0002-0590-1022}{\includegraphics[scale=0.06]{orcid.pdf}\hspace{1mm}Daniel T. Welling} \\
	Department of Climate and Space\\
	University of Michigan\\
	Ann Arbor, MI\\
	\\
 	\And
	\href{https://orcid.org/0000-0003-4515-0208}{\includegraphics[scale=0.06]{orcid.pdf}\hspace{1mm}Brian A. Larsen} \\
	Space Science and Applications\\
	Los Alamos National Laboratory\\
	Los Alamos, NM\\
	\\
}

\date{}

\renewcommand{\shorttitle}{SpacePy at 12}

\hypersetup{
pdftitle={The SpacePy Space Science Package at 12 Years},
pdfsubject={space-physics},
pdfauthor={J.T. Niehof, S.K. Morley, D.T. Welling and B.A. Larsen},
pdfkeywords={python, software, spacepy, cdf, datamodel, coordinates, time, magnetosphere, heliosphere, open source},
}

\begin{document}
\maketitle

\begin{abstract}
	For over a decade, the SpacePy project has contributed open-source solutions for the production and analysis of heliophysics data and simulation results. Here we introduce SpacePy's functionality for the scientific user and present relevant design principles. We examine recent advances and the future of SpacePy in the broader scientific Python ecosystem, concluding with some of the work that has used SpacePy.
\end{abstract}


\section{Introduction}

The roots of the SpacePy project \citep{SpacePy} date back to 2009, although the first public presentation of SpacePy was to the 9th Python in Science conference in 2010 \citep{spacepy2011}. The mission statement of the open source library, given in both the conference proceedings and on the original web page, was ``to promote accurate and open research standards by providing an open environment for code development. In the space physics community there has long been a significant reliance on proprietary languages that restrict free transfer of data and reproducibility of results. By providing a comprehensive library of widely-used analysis and visualization tools in a free, modern and intuitive language, we hope that this reliance will be diminished for non-commercial users.'' \citep{spacepy2011}

Now, twelve years after the presentation of SpacePy and eleven after its initial open source release, we present a summary of the present state of the library, a retrospective view of SpacePy's development, and a look to the future of SpacePy and its place in the heliophysics scientific software ecosystem \citep[e.g.,][]{Burrell2018}. For scientists in the field, we introduce the research-enabling functionality of SpacePy and the scientific Python ecosystem, including examples of previous studies. For research software engineers \citep[e.g.,][]{crouch2013}, we discuss how SpacePy is designed to interoperate with the greater technical and social ecosystem of heliophysics software.

The domain of SpacePy is space physics broadly speaking, i.e. heliospheric and magnetospheric physics, including magnetosphere-ionosphere coupling. Strict solar physics and isolated ionospheric physics are outside of the usual scope, although SpacePy functionality may be useful in those fields.

\section{Design Goals}
SpacePy is designed as a library; that is, its primary access is via the public application programming interface (API), rather than a user-facing application. Target users are scientists and engineers writing custom code for specialized analysis and visualization of data or model results, producing archival-quality data sets, or creating a more user-facing interactive application. The developers consider SpacePy a success when it is used to provide functionality to higher-level codes and uses lower-level libraries to provide the ``building blocks'' of such functionality.

As with other scientific libraries, the reliability and fidelity of results is essential. A thorough testing suite ensures reproducibility of results, and a test-first approach to fixing bugs prevents regressions. Absolute accuracy of results up to the numerical precision of the computer is considered less important than documenting the expected precision, the regimes in which results are reliable, and the source of the algorithm, including citation of the literature where appropriate (e.g. in empirical models).

SpacePy development is user-driven: functionality is developed to meet a specific scientific or mission operations goal. Developers are scientists in the field. This ensures applicability of the implementation; the distinction between SpacePy and project-specific code is the conversion to maintained, tested, and widely applicable functionality.

On the computational side, the API aims to be carefully designed, ``Pythonic'' in nature \citep{Alexandru2018} and in accordance with software engineering good practices such as abstraction. One illustration of the success of this approach is the pycdf interface (section~\ref{sec:cdf}). Although independently developed, the resulting interface is very similar to the h5py HDF5 library \citep{collette_python_hdf5_2014}; the SpacePy datamodel was developed along similar lines at the same time. This minimizes the cognitive load required to access similar data from differing container formats, allowing the user to focus on problem solving rather than interface peculiarities.

SpacePy has been available to the general public under an open source license since 2011. The SpacePy license is essentially that of Python itself, with the only change being replacing references to Python with SpacePy, and to the Python Software Foundation with Triad National Security, LLC, as the initial licensor. This license, often called ``the PSF license'', is a BSD-style permissive license approved by the Open Source Initiative, the Free Software Foundation, and the Debian Free Software Guidelines. Before arrangements were made for this public release, SpacePy was briefly provided under a restrictive noncommercial license upon request \citep{spacepy2011}.

The SpacePy install and update process supports a range of deployment and update strategies. Although the most common means of installation is automatic management via \texttt{pip}, manual download and installation from source remain options for those users who wish to install into a shared location on a multi-user system, do not have full Internet access on their deployed system, or have other particular needs. Similarly, SpacePy supports a wide range of versions of its dependencies and changes these requirements at specific version numbers only (where the subminor version is 0, e.g. 0.2.0, 0.3.0). API changes are also made at predetermined version numbers, with deprecation warnings providing a graceful migration path. The documentation clearly states versions of API changes, even for versions in the past, to support users updating their code. The SpacePy team recognizes that users have a range of needs, may have limited control of their operating environment, and need to interoperate with other packages which may have stricter requirements; thus SpacePy is designed to be as flexible as practicable on these issues.

SpacePy takes a balanced approach to using other packages as dependencies: maximizing the use of mature, robust dependencies decreases the maintenance load of SpacePy itself, as well as enhancing interoperability, but may place additional burden on users (even those who do not use the functionality for which a dependency is required). The approach is to bring in a dependency where it provides significant (rather than incidental) functionality, ideally supporting multiple components of SpacePy. The specifics are left intentionally vague. More importantly, the functionality provided by each dependency is explicitly documented, and SpacePy will install without most dependencies. Section~\ref{sec:future} describes the future direction of dependency handling.

Over the past few years (section~\ref{sec:recent}), compliance with Python in Heliophysics Community (PyHC) standards \citep{annex_a_2018_2529131} has been a major design consideration.

\section{Capability and Architecture}
\begin{figure}[h!]
\begin{center}
\includegraphics[width=10cm]{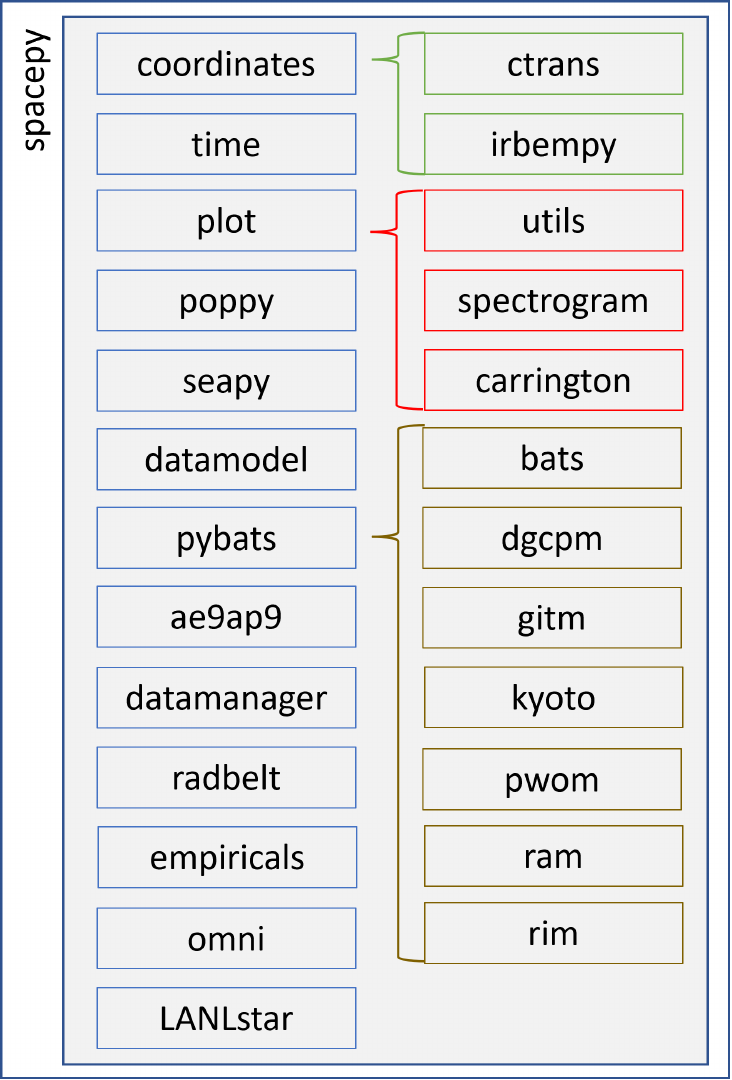}
\end{center}
\caption{Core capabilities organized by namespace in the SpacePy package.}\label{fig:1}
\end{figure}

All capabilities described in this section are available in the current release, SpacePy 0.4.0, available at \url{https://pypi.org/project/spacepy/}. Capabilities are also summarized in the SpacePy documentation at \url{https://spacepy.github.io/capabilities.html}.
A graphical overview of key namespaces (i.e., modules) in SpacePy is shown in Figure~\ref{fig:1}.

\subsection{Datamodel}
\label{sec:datamodel}
On of the core capabilities of SpacePy is its data model representation, which was introduced shortly after \citet{spacepy2011}.
SpacePy uses a description, based on that used by HDF5, which uses three key concepts: groups, datasets, and attributes.
Groups are analogous to file system directories, and can contain both groups and datasets. Datasets are n-dimensional arrays of data.
Attributes are metadata that is carried with either a group or a dataset.
SpacePy's \texttt{spacepy.datamodel.SpaceData} class implements the \textit{group} by subclassing \texttt{dict}, while the \texttt{spacepy.datamodel.dmarray} implements the \textit{dataset} as a subclass of \texttt{numpy.ndarray}.
Each of these classes carries metadata in a Python dictionary accessed via the \texttt{attrs} attribute.
The structure of the object can be displayed using the \texttt{spacepy.datamodel.SpaceData.tree} method.

\begin{verbatim}
>>> import spacepy.datamodel as dm
>>> sdata = dm.SpaceData(attrs={"Descriptor": "Test file"})
>>> sdata["Variable0"] = dm.dmarray([0,1,2],
                                    attrs={"CATDESC": "An increasing value"})
>>> sdata["Variable1"] = dm.dmarray([2.7]*3,
                                    attrs={"CATDESC": "The value 2.7 repeated"})
>>> sdata.tree(attrs=True)
+
:|____Descriptor
|____Variable0
    :|____CATDESC
|____Variable1
    :|____CATDESC
\end{verbatim}

The datamodel thus provides a file-format agnostic representation of data that preserves metadata. The data model objects can be constructed and used without requiring either input or output, however, read and write support is provided. Supported file formats include:

\begin{itemize}
    \item NASA CDF: NASA's Common Data Format
    \item HDF5: Hierarchical Data Format 5
    \item NetCDF: Unidata's Network Common Data Form
    \item JSON-headed ASCII
\end{itemize}

Spacepy's \texttt{datamodel} readers are currently all \textit{greedy} by default, in that they load files all-at-once. While this is convenient for many users, for very large data files or for systems with read/write speed limitations this can be sub-optimal.

The datamodel is normally agnostic to the interpretation of metadata, so it can be used for a wide range of metadata standards. This may be a simple human-readable informal representation. Additional functions are provided for the case where metadata are ISTP/SPDF compliant.

\subsubsection{NASA's Common Data Format}
\label{sec:cdf}
SpacePy has provided first class support for NASA CDF, including full read and write, since September 2010 through the \texttt{spacepy.pycdf} module. \texttt{pycdf} provides a pythonic interface to the NASA CDF library, and requires that the user obtain that library from NASA.
This approach is taken to reduce duplication of functionality and maintain a clear separation of responsibility: NASA develops and maintains CDF, while SpacePy develops and maintains the Python interface. pycdf reads files ``on demand'', with the ability to read a single variable or even fraction thereof. \texttt{spacepy.datamodel.fromCDF} provides an at-once read into the SpacePy datamodel.

\subsubsection{HDF5 and derivatives}
Several other formats and packages derive from HDF5 and can, unless non-standard features are added, be read directly with \texttt{spacepy.datamodel.fromHDF5}. For example, since MATLAB$^{\textrm{\textregistered}}$ release R2006b, \texttt{.mat} files can be (and are most likely to be) stored as HDF5 files. Also, NetCDF4 provides an alternative API to build and read data files using the HDF5 library. NetCDF4 files can thus be read using \texttt{spacepy.datamodel.fromHDF5}. Note that NetCDF3 is not compatible with HDF5, even though NetCDF4 provides access to legacy NetCDF3 files. \texttt{spacepy.datamodel.fromNC3} provides NetCDF-to-datamodel reader functionality by building on the \texttt{scipy.io.netcdf} reader.

To write the contents of a \texttt{spacepy.datamodel.SpaceData} to an HDF5 file, simply call the appropriate write method:
\begin{verbatim}
>>> sdata.toHDF5("output_filename.h5")
\end{verbatim}

\subsubsection{JSON-headed ASCII}
This is a text-based data format that uses a header, written in JavaScript Object Notation (JSON) and intended to be both human- and machine-readable, to describe the file layout and to store metadata.
While not in broad use, this provides specific support for the magnetic ephemeris (``magephem'') files for the Van Allen Probes Energetic particle, Composition, and Thermal plasma (RBSP-ECT) Suite, as well as the energetic particle data from the Global Positioning System \citep{morley2017}. This format is also supported by \textit{Autoplot} \citep{faden2010}

\subsection{Time Systems}
Handling time and coordinate systems is fundamental to much of space physics.
While these capabilities were present in the original release of SpacePy, there have been significant advances over the years.

\begin{figure}[h!]
\begin{center}
\includegraphics[width=10cm]{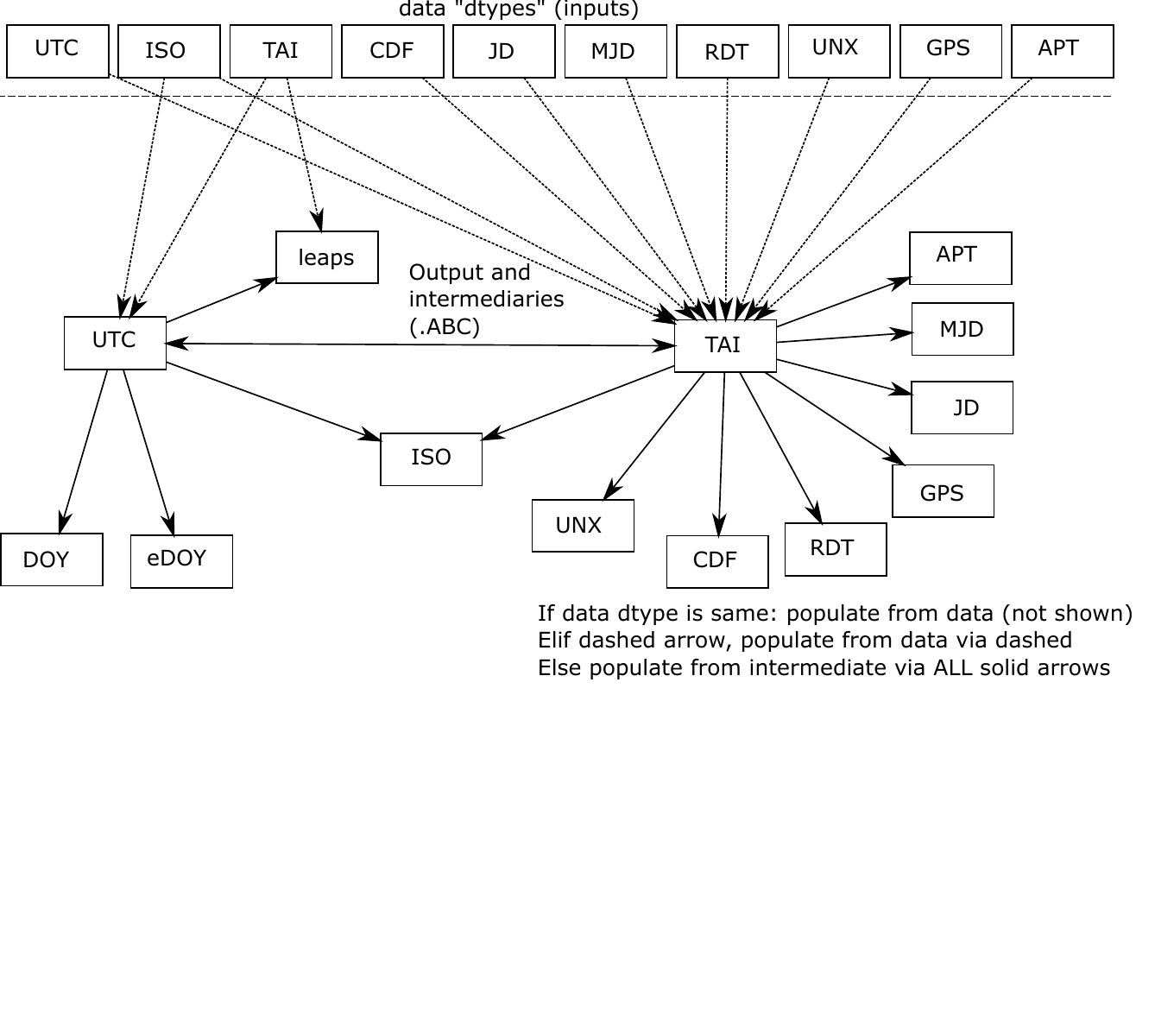}
\end{center}
\caption{Workflow for supported input and output types in \texttt{spacepy.time}}\label{fig:2}
\end{figure}

SpacePy supports multiple time systems: Coordinated Universal Time (both as native Python \texttt{datetime} objects, and expressed as ISO8601 time strings); International Atomic Time (TAI), in seconds since 1958-01-01T00:00:00\,UTC; Global Positioning System (GPS) time, in seconds since 1980-01-06T00:00:00\,UTC; Julian Day and Modified Julian Day (expressed on the UTC scale); Unix time; Rata Die time, in days since 0001-01-01T00:00:00\,UTC; and CDF time (corresponding to the legacy \texttt{CDF\_EPOCH} types in NASA's CDF library). Figure~\ref{fig:2} shows relationships between these time systems; internal processing is primarily in TAI.

\subsubsection{Handling Leap Seconds}
Some time systems ignore leap seconds (e.g. Unix time).
Similarly, many standard library time packages do not handle leap seconds, including Python's \texttt{datetime} module (even as used for UTC).
On the other hand, there is a need in heliophysics to represent leap seconds and to convert between continuous time representations and those that ignore leap seconds. These conversions are well-defined from the introduction of leap seconds to UTC in 1972 to the present.
For systems that cannot represent leap seconds, the leap second moment is considered not to exist. For example, from 23:59:59 on 2008-12-31 to 00:00:00 on 2009-01-01 is two seconds, but only represents a one-second increment in Unix time.

SpacePy uses an user-updatable leap second table referenced to the latest US Naval Observatory data.

\subsection{Coordinates}
Since its first release SpacePy has provided a pythonic interface to the IRBEM library \citep{irbem}. This includes access to magnetic field models, field line tracing, and coordinate transformations.
In release 0.3 SpacePy introduced a new backend for handling coordinate system transformations, while simultaneously preserving the familiar \texttt{spacepy.coordinates.Coords} interface.
This new backend maintained existing functionality, requiring no changes to existing code, while removing the need for Fortran support (for the IRBEM library) to perform coordinate transformations. Both backends are available to the user.

\begin{figure}[h!]
\begin{center}
\includegraphics[width=10cm]{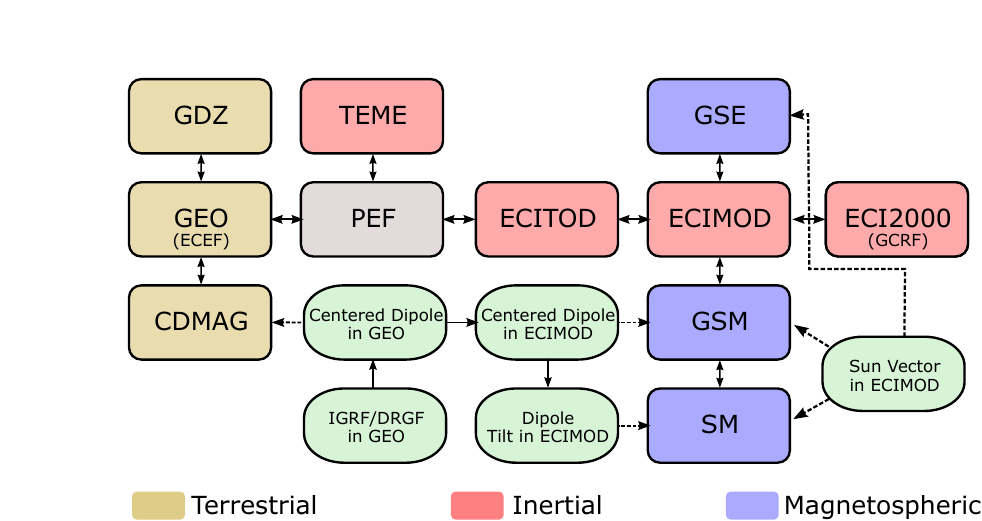}
\end{center}
\caption{Relationships between coordinate systems as implemented in \texttt{spacepy.ctrans}}\label{fig:3}
\end{figure}

Coordinate systems supported by this module broadly fall into two categories: those that can be defined strictly using astronomical parameters only, and those that require a representation of Earth's geomagnetic field.
SpacePy uses transformations that build from the IAU 1976/FK5 system for astrophysical reduction \citep{lederle1980,fricke1982,seago2000}. Taking the origin of our coordinate systems as the center of the Earth instead of the solar barycenter gives us an Earth-centered inertial (ECI) system as our starting point.
The relationships between the supported coordinate systems are described below, and graphically summarized in Figure~\ref{fig:3}.
In contrast with many other space or heliophysics packages we do not follow the approach given by \citet{russell1971} or \citet{hapgood1992} of using first order approximations to the reduction theory. SpacePy uses the full third order relationships in its implementation.

\subsubsection{Earth-Centered Inertial Systems}
Our fundamental reference system is \texttt{ECI2000}, sometimes simply referred to as the J2000 frame, though we avoid this to prevent confusion with the J2000 epoch (January 1$^{st}$ 2000, 11:58:55.816 UTC)
This system can be considered equivalent to the Geocentric Celestial Reference Frame, to within tens of milliarcseconds.The z-axis is perpendicular to the mean celestial equator at the J2000 epoch. The x-axis is aligned with the mean equinox at the J2000 epoch. The y-axis completes and lies in the celestial equatorial plane.

Correcting the orientations of the equator and equinox for precession yields the mean equinox and mean equator of date, and updating the definition gives us \texttt{ECIMOD} (ECI Mean Of Date).
Finally, we account for the nutation (the short- period perturbations on the precession) to obtain the true equator and true equinox of date. Using these corrected axes to define our ECI system gives \texttt{ECITOD} (ECI True Of Date).

\subsubsection{Terrestrial systems: Geographic, Geodetic, and Geomagnetic}
SpacePy implements an Earth-Centered Earth-Fixed coordinate system using the name \texttt{GEO} (Geocentric Geographic). The coordinates of a point fixed on (or relative to) the surface of the Earth do not change as the Earth rotates. The x-axis lies in the Earth’s equatorial plane (zero latitude) and intersects the Prime Meridian (zero longitude; Greenwich, UK). The z-axis points to True North (which is roughly aligned with the instantaneous rotation axis).

While all of the coordinate systems thus far are generally defined as Cartesian systems, geodetic (\texttt{GDZ}) coordinates cannot be properly represented as Cartesian. \texttt{GDZ} is defined in terms of altitude above a reference ellipsoid, the geodetic latitude, and geodetic longitude. Geodetic longitude is identical to geographic longitude, while both the altitude and latitude depend on the ellipsoid used. SpacePy's default is the WGS84 reference ellipsoid and the \texttt{GEO}-\texttt{GDZ} conversion uses Heikkinen's exact algorithm \citep[see][]{zhu1994}.

Finally, geomagnetic coordinates can be considered a magnetic analog of \texttt{GEO}. The z-axis is aligned with the centered dipole axis of date (defined using the first 3 coefficients of the IGRF/DGRF). The y-axis is perpendicular to to both the dipole axis and True North and the x-axis completes the system.

\subsubsection{Magnetospheric Systems}
Magnetospheric coordinate systems are non-inertial and Earth-centered. We begin with \texttt{GSE} (Geocentric Solar Ecliptic). The x-axis points from the center of Earth to the solar system barycenter, while the y-axis is defined to lie in the mean ecliptic plane of date (pointing in the anti-orbit direction) and the z-axis is perpendicular to the mean ecliptic plane.

To move to \texttt{GSM} (Geocentric Solar Magnetospheric) we  require that the centered dipole axis lies in the x-z plane. The y-axis is thus perpendicular to both the Sun-Earth line and the centered dipole axis . GSM is therefore a rotation about the x-axis from the GSE system. Finally, we move to \texttt{SM} (Solar Magnetic) where the z-axis is aligned with the centered dipole axis of date (positive northward), and the y-axis is perpendicular to both the Sun-Earth line and the dipole axis. As with \texttt{GSE} and \texttt{GSM}, y is positive in the anti-orbit direction. The x-axis therefore is not aligned with the Sun-Earth line and \texttt{SM} is a rotation about the y-axis from the \texttt{GSM} system.

We note that these definitions differ slightly from those used by, e.g., \citet{hapgood1992} as the mean ecliptic (correcting for precession) is used instead of the true ecliptic (correcting for precession and nutation), with the Earth-Sun vector also defined in \texttt{ECIMOD}. However, they have been adopted for consistency with recent flagship missions following the implementations used for Van Allen Probes and Magnetospheric Multiscale \citep[e.g.,][]{morley_steven_karl_2015_2594027}.

\subsection{Pybats} 

The Pybats module of Spacepy provides tools for handling output from the Space Weather Modeling Framework \citep{toth2005,Toth2012,Gombosi2021}.
The SWMF is a framework that executes, synchronizes, and couples together many physics-based domain models of the complex heliosphere system, from solar corona to planetary atmospheres \citep[e.g.,][]{Powell1999,Welling2015,Mukhopadhyay2020,Sachdeva_2021}. 
It is widely used in heliophysics, including long-standing availability at NASA's Community Coordinated Modeling Center (CCMC) and real-time operational use at NOAA's Space Weather Prediction Center (SWPC) since 2016.
Its wide adoption has necessitated a tool box for reading and handling its complex and heterogeneous output -- a need met by the Pybats module.

The fundamental goal of Pybats is to allow users to access SWMF model output within Python environments.
It achieves this by subclassing \texttt{spacepy.datamodel.SpaceData} to include file read methods called upon instantiation. 
This allows for easy exploration of values and attributes as outlined above.
In the base \texttt{spacepy.pybats} module, classes are provided for data formats defined at the SWMF control level or common across many SWMF sub-models.
Sub-modules provide model-specific functionality and customization of base classes.
For example, the BATS-R-US global MHD model \citep{Powell1999, DeZeeuw2000, Groth2000} produces basic ASCII log files that follow a standard SWMF-defined format and are readable via \texttt{spacepy.pybats.LogFile} objects.
However, the \texttt{spacepy.pybats.bats} submodule provides model-specific classes and capabilities.
When opening BATS-R-US log files, the \texttt{spacepy.pybats.bats.BatsLog} subclass includes additional methods for visualizing values inherent to the BATS-R-US output data, such as Dst index.
Conversely, the output files from the Ridley Ionosphere Model (RIM, \cite{Ridley2001}) are proprietary formats, so the base classes for handling RIM output are located in the \texttt{spacepy.pybats.rim} sub-module.

The most fundamental file type handled by Pybats is the SWMF IDL format, which has suffix \texttt{\*.out}.
These files are of a format proprietary to the SWMF, may be either ASCII or binary, and can hold 1, 2, or 3D data sets.
These files can be concatenated together to hold multiple epochs of simulation of data in a single file (a \texttt{\*.outs} file), allowing users to reduce the total number of files produced by a single simulation.
SWMF IDL files are used by many different models, including BATS-R-US, PWOM, DGCPM, and others.
The base class \texttt{spacepy.pybats.IdlFile} automatically detects file format (ASCII versus binary) upon instantiation, reads the file into a \texttt{spacepy.datamodel.SpaceData}-like object, and provides tools to navigate the different frames, or single-epoch sets, stored within the file.

An animation of SWMF output using SpacePy is available in the supplemental material and at \url{https://www.youtube.com/watch?v=8bgkgQITFO8}. This animation shows the magnetospheric response as the interplanetary magnetic field switches from a purely northward to purely southward direction. The simulation was performed using the SWMF, coupling the BATS-R-US global MHD model with the Rice Convection Model and the RIM. For this simulation, the physics-based Adaptive Mesh Refinement (AMR) capability of BATS-R-US was used to automatically increase spatial resolution to a minimum of 1/8 Earth Radii (R$_E$). The grid was refined in any block where the current density surpassed $10^{-5} \mu A/m^2$ and coarsened if the current dropped below $5\times 10^{-7} \mu A/m^2$.
Visualization of the model output was performed entirely with SpacePy's pybats module and submodules.
Current density contours in the equatorial plane were plotted using the \texttt{spacepy.pybats.Bats2d.add\_contour} method. The colored squares show the BATS-R-US block tree structure; the color of each square shows the grid resolution of the block with brighter colors indicating the regions of finest grid spacing. The \texttt{spacepy.pybats.Bats2d.add\_grid\_plot} method was used to add the grid information to screen. The animation demonstrates how BATS-R-US AMR can be used to simulate fine structure within the magnetosphere, including Kelvin-Helmholtz instabilities, flux transfer events, and fast flow channels in the tail.

\subsection{Interoperability}
To maximize flexibility for the researcher and minimize duplication of effort, SpacePy emphasizes interoperability with other packages. SpacePy's reliance on the widely-used NumPy \citep{harris2020array} array package provides a baseline of low-level interoperability, and the datamodel (section~\ref{sec:datamodel}) was designed to make the minimum changes necessary to the NumPy array interface.

SpacePy's \texttt{Ticktock} time object supports conversion to and from Astropy's \citep{2013A&A...558A..33A,2018AJ....156..123A} \texttt{astropy.time.Time} representation; similarly, SpacePy \texttt{Coords} can be converted to and from the Astropy \texttt{astropy.coordinates.SkyCoord}. Both conversions are via simple to/from methods of the SpacePy objects.

\texttt{SkyCoord} conversion is performed via the Earth-centered Earth-fixed frame (\texttt{GEO} in SpacePy, \texttt{ITRS} in astropy). \texttt{Time} conversion uses SpacePy's TAI format and Astropy's TAI scale with GPS format, both being continuously-running counts of seconds since a defined epoch.

Transformation of data structures to and from additional packages is in preparation (section~\ref{sec:recent}).

\subsection{Empirical Models}
Via the Pythonic interface to the IRBEM library (\texttt{irbempy}), SpacePy supports a wide range of magnetic field models and operations on them, including field line and drift shell tracing. The LANL* neural net based model \citep{2012SpWea..10.2014Y,Yu2014} provides faster calculation of the third adiabatic invariant and the last closed drift shell. This model has recently been migrated from the Fortran-based ffnet library to a new implementation based on numpy linear algebra routines, while maintaining the neural network structure, weights, and results.

Other empirical models include plasmapause models, the magnetopause model of \cite{1997JGR...102.9497S}, and access to the output of the AE9/AP9 radiation belt model \citep{2013SSRv..179..579G}. As inputs to these and other models, SpacePy provides the \texttt{omni} module, simplifying access to the upstream solar wind data set of \cite{omni} using the interpolation techniques of \cite{qin}.

\section{Recent Activities}
\label{sec:recent}

In the summer of 2018, SpacePy transitioned from an open source release model to a fully open development model. All development is done in a ``live'' github repository at \url{https://github.com/spacepy/spacepy}, issues and enhancements are processed with full public visibility, and developer commits go through the same review and workflow as outside contributors. The result has been not only feature requests and issues from the user community, but also new and improved functionality. AstroPy coordinate support and the new LANL* processing are two examples where the core functionality came from the community and was integrated into SpacePy with developer support.

The transition away from Python 2 is concluding. Although SpacePy has fully supported Python 3 since version 0.1.5 (December 2014), Python 2 support was retained. This has been slowly phased out over several releases, providing time for users to update. Soon Python 2 code will be removed, simplifying the codebase and facilitating further transitions, such as the move away from distutils, which were not possible while supporting Python 2. We successfully supported a dual-version codebase with very little version-specific code for over seven years.

An ongoing project will connect the datamodel of SpacePy with the HAPI streaming Heliophysics data protocol \citep{2021JGRA..12629534W} and the data structures of the SunPy library \citep{sunpy_community2020}. This work supports the use of functionality in a range of libraries without forcing users into a single data representation or a single library ecosystem. We intend continual interoperability with other packages within the broader scientific Python community.

Part of the datamodel conversions project is extending the ability of SpacePy to interpret ISTP/SPDF metadata \citep{ISTP} regardless of its container. Ultimately this will allow the easy manipulation of data using the ISTP metadata standard regardless of its container (SpaceData, HDF5, or CDF). This will not change the fundamental nature of the SpacePy datamodel, which is agnostic to the form of metadata, only allow additional functionality where the metadata are ISTP-compliant.

SpacePy developers have been regularly engaging with the PyHC, including participation in the 2022 Summer School (\url{https://heliopython.org/summer-school}).
\section{Applying SpacePy}
SpacePy has been used in many scientific studies as well as in support of mission data processing; only a few examples are provided here.

Recent uses of SpacePy in scientific publications range from probabilistic predictions of geomagnetic storms \citep{2020JSWSC..10...36C}, visualization and verification of an improved inner magnetosphere model \citep{Engel2019}, through calculation of L-shells on Cubesats \citep{2020AdSpR..66...52G} to modeling of geomagnetic response to a ``perfect storm'' ICME \citep{2021SpWea..1902489W}.

In missions, SpacePy supported the data processing for the Radiation Belt Storm Probes Energetic particle, Composition, and Thermal plasma suite (RBSP-ECT) \citep{2013SSRv..179..311S,VanAllenSOCs}, including the ECT combined electron product \citep{2019JGRA..124.9124B}. SpacePy supports data management within the Magnetospheric Multiscale mission (MMS) magnetic ephemeris processing chain \citep{morley_steven_karl_2015_2594027}. Data from the Integrated Science Investigation of the Sun suite \citep{2016SSRv..204..187M} on Parker Solar Probe are processed with SpacePy.

Functionality used in earlier studies remains fully maintained and available for other studies, such as superposed epoch analysis \citep{2010RSPSA.466.3329M,rogers_2022} and association of point processes \citep{2012AnGeo..30.1633N}.

The SpacePy team maintains a list of publications at \url{https://spacepy.github.io/publications.html} and welcomes submissions.

\section{Future Directions and Challenges}
\label{sec:future}
It is clear that the Heliophysics community move away from IDL is well underway, so the SpacePy goal of reducing ``reliance on proprietary languages'' is at least partially accomplished, through the efforts of many in the community. Proprietary languages are likely to retain some importance but the place of Python as a tool is well established. One significant question then is what the nature of the Python in Heliophysics ecosystem will be.

Since Python is an easy language to write, and modern environments such as github and the Python Package Index (PyPI) make sharing easy, Heliophysics-related Python packages have proliferated. This has resulted in potential issues of duplication of effort and interoperability between packages. The PyHC project has done an excellent job of making packages aware of each other so that they can voluntarily evaluate existing functionality, avoid duplication, and work on interoperability. Given diversity of workflows, facilitating this work is more likely to be successful than any attempt to force the community into a single approved package for each function. Interoperability does raise the possibility of circular dependencies, but this need not be a problem. As long as packages do not depend on each other for \textit{installation}, modern package managers will successfully install both. Careful interface design can then avoid circular imports; this has been the case for the datamodel interoperability project (which will also produce a set of recommendations for facilitating interoperability).

As the scientific Python ecosystem grows, SpacePy's dependency strategy is constantly evolving. One solution may be for some generic SpacePy functionality to migrate ``up the stack'' into e.g. scipy; another (not exclusive) may be to use the optional specifications of PEP508 \citep{pep508} to only install dependencies for SpacePy functionality that a user specifically requests.

One major shift over the life of SpacePy has been the transition from source-based distribution to binary-based (e.g. operating-system specific binary wheels). This places additional demands on package developers, not only in producing these binaries but in supporting newer build systems. The result can be a substantial improvement in ease of installation for the end user, and SpacePy is transitioning away from the assumption that a user will have a working compiler, even on Unix-based systems. Maintaining flexibility of deployment remains a priority. Supporting this installer transition requires significant computer engineering work which is largely separate from the domain expertise.

To date, SpacePy development has been supported primarily via the missions that benefit from it. Short-term independent support has been secured to support engineers in addressing these computer engineering based tasks more efficiently than using physics domain experts. We hope similar support will continue across the Python ecosystem, as it is essential to high-quality software.

\section*{Conflict of Interest Statement}

The authors declare that the research was conducted in the absence of any commercial or financial relationships that could be construed as a potential conflict of interest.

\section*{Author Contributions}

SM, JN, and DW wrote the initial draft and made substantial revisions and additions to all portions of the manuscript. 
All authors contributed to the manuscript concept, read and approved the submitted version, and have routinely contributed code, documentation, and reviews to SpacePy over the past decade.

\section*{Funding}
SpacePy development has been supported by several missions, including the Van Allen Probes Radiation Belt Storm Probe, Energetic particle, Composition, and Thermal plasma suite, JHU/APL contract 967399 under NASA prime contract NAS5-01072; and the Parker Solar Probe Integrated Science Investigation of the Sun, JHU/APL contract 136435 under NASA prime contract NNN06AA01C. Further support was provided by U.S. Government contract 89233218CNA000001 for Los Alamos National Laboratory (LANL), which is operated by Triad National Security, LLC for the U.S. Department of Energy; and NASA grant 80NSSC21K0304 ``Enhancing Heliophysics Python Library Interoperability by Adapting Common Data Models''.

\section*{Acknowledgments}
The authors would like to thank everybody who has contributed to SpacePy via pull requests, discussions, and well-formed issues.

\section*{Data Availability Statement}

The reference of record for SpacePy code is \cite{SpacePy}. SpacePy users are also encouraged to cite the present work in studies which make use of SpacePy. Code releases are available via the PyPI at \url{https://pypi.org/project/spacepy/}, development is hosted at \url{https://github.com/spacepy/spacepy/}, and documentation at \url{https://spacepy.github.io/}.


\section*{Figure captions}

\end{document}